\documentclass[letterpaper]{article} 
\usepackage[submission]{aaai23}  
\usepackage{times}  
\usepackage{helvet}  
\usepackage{courier}  
\usepackage[hyphens]{url}  
\usepackage{graphicx} 
\usepackage{natbib}  
\usepackage{caption} 
\usepackage{algorithm}
\usepackage{algorithmic}
\usepackage{booktabs}
\usepackage{newfloat}
\usepackage{listings}
\DeclareCaptionStyle{ruled}{labelfont=normalfont,labelsep=colon,strut=off} 
\floatstyle{ruled}
\newfloat{listing}{tb}{lst}{}
\floatname{listing}{Listing}
\title{Exploring the Efficacy of Pre-trained Checkpoints \\ in Text-to-Music Generation Task}
\author {
    Shangda Wu,$^{\dagger}$
    Maosong Sun$^{\dagger\ddagger}$
}
\affiliations {
    $^{\dagger}$ Department of Music AI and Information Technology, Central Conservatory of Music\\
    $^{\ddagger}$ Department of Computer Science and Technology, Tsinghua University\\
    shangda@mail.ccom.edu.cn, sms@tsinghua.edu.cn
}

\usepackage{bibentry}

\begin{document}

\maketitle

\begin{abstract}
Benefiting from large-scale datasets and pre-trained models, the field of generative models has recently gained significant momentum. However, most datasets for symbolic music are very small, which potentially limits the performance of data-driven multimodal models. An intuitive solution to this problem is to leverage pre-trained models from other modalities (e.g., natural language) to improve the performance of symbolic music-related multimodal tasks. In this paper, we carry out the first study of generating complete and semantically consistent symbolic music scores from text descriptions, and explore the efficacy of using publicly available checkpoints (i.e., BERT, GPT-2, and BART) for natural language processing in the task of text-to-music generation. Our experimental results show that the improvement from using pre-trained checkpoints is statistically significant in terms of BLEU score and edit distance similarity. We analyse the capabilities and limitations of our model to better understand the potential of language-music models.
\end{abstract}

\section{Introduction}
Creativity was once thought to be a privilege of humans, but after training on large amounts of data, transformer-based models \cite{DBLP:conf/nips/VaswaniSPUJGKP17} also exhibit this capability to some extent. Recent transformer-based models can generate human-like texts \cite{DBLP:conf/nips/BrownMRSKDNSSAA20}, autocomplete codes \cite{DBLP:journals/corr/abs-2107-03374}, reconstruct images \cite{DBLP:conf/cvpr/HeCXLDG22}, or compose music \cite{DBLP:journals/corr/abs-2005-00341}. Models designed by researchers focused on AI creativity have also shown mind-blowing generation results across modalities. For example, DALL·E 2 \cite{DBLP:journals/corr/abs-2204-06125}, a text-conditional image generation model, can generate realistic images and creative art from natural language captions. On the other hand, AudioGen \cite{DBLP:journals/corr/abs-2209-15352}, a textually guided audio generation model, although trained on general audio, can also generate music clips by giving proper textual descriptions.

Symbolic music, unlike raw audio, contains explicit musical information, such as note onsets and pitch on individual tracks. With the development of deep learning technology, symbolic music generation \cite{DBLP:journals/nca/Briot21} has shown unprecedented progress in the past few years. However, the lack of datasets has been a major limitation in the task of symbolic music generation. Due to the absence of text-music pairs data, the task of text-conditional symbolic music generation has not been given enough attention in the past.

Even without text-music datasets, a few researchers managed to achieve the conversion of input text into symbolic music. Rangarajan designed three strategies \cite{DBLP:conf/icdim/Rangarajan15} for mapping text to notes. But as it is based on character-level mappings, the generated music is very random and does not reflect the semantic information in the text. TransProse \cite{DBLP:conf/clfl/DavisM14} contains several mapping rules that generate music based on the density of emotional words \cite{DBLP:journals/ci/MohammadT13} in the given text. However, TransProse does not reflect the non-emotional information in the text, and its creativity is limited by those hand-crafted mapping rules. BUTTER \cite{zhang2020butter}, a GRU-based model, can search and generate music segments given rigid text descriptions and vice versa. The dataset used by BUTTER contains 16,257 folk songs, and the paired text descriptions are synthesised from keywords (i.e., 25 keys, 6 meters, and 3 styles). Although BUTTER is a data-driven model, its flexibility is limited as the paired texts are synthesised from specified keywords, and it can only generate 16-beat (4-bar) music segments. The most recent attempt is Mubert\footnote{https://github.com/MubertAI/Mubert-Text-to-Music}, which can generate music from user-given prompts. Although its generated music is of high quality, it does not directly generate music from the input text. The input text and Mubert API tags are encoded by Sentence-BERT \cite{DBLP:conf/emnlp/ReimersG19}, and the closest tag vectors are selected and then used for music generation. All sounds are created beforehand by musicians and sound designers, and thus Mubert is more like generating a combination of sounds, instead of music.

In this paper, we model the task of text-to-music generation as a sequence-to-sequence problem, and develop a transformer-based model that is capable of generating complete and semantically consistent music scores directly from descriptions in natural language based on text\footnote{https://github.com/sander-wood/text-to-music}. To the best of our knowledge, this is the first model that achieves text-conditional symbolic music generation which is trained on real text-music pairs, and the music is generated entirely by the model without any hand-crafted rules. We further explore the efficacy of using publicly available pre-trained BERT, GPT-2, and BART checkpoints, and aim to provide empirical answers to the following research questions.

\begin{itemize}
    \item Does using pre-trained checkpoints improve the performance of language-music models?
    \item To what extent can language-music models learn the relationship between natural language and symbolic music?
\end{itemize}
\vspace{-1em}
\textbf{\textit{}} 

We believe that researchers from the fields of symbolic music generation and natural language processing can take insights from this paper when dealing with language-music problems in the future.

\section{Dataset}
Large-scale data is the cornerstone that underpins data-driven models. For example, recent multimodal language-vision models are usually trained on hundreds of millions of image-text pairs \cite{DBLP:journals/corr/abs-2111-02114}. Likewise, large-scale paired text-music datasets are a prerequisite for language-music models.

To make models learn the relationship between natural language and symbolic music, we collected as many text-music pairs as possible, composing a text-music dataset called \textit{Textune}. This dataset contains 282,870 English text-tune pairs, where all tunes are represented in ABC notation\footnote{https://abcnotation.com/}. As ABC notation encodes music scores into sequences of ASCII characters, using this music notation system makes it easy to model the text-to-music generation task as a sequence-to-sequence problem.

All scores in Textune can be written on one stave (for vocal solo or instrumental solo) in standard classical notation, and are in a variety of styles, e.g., blues, classical, folk, jazz, pop, and world music. The scores are not all western, but also include some from other regions (e.g., Asia and Africa). In addition, all the scores have at least eight bars to present a complete musical idea.

Due to the abstract nature of music, it is much more difficult to describe music accurately than to describe images. The description of the same piece of music can vary significantly from person to person according to their respective musical backgrounds. In general, the valid text descriptions in Textune can be categorised as follows: 1) musical analysis (e.g., tonal analysis and harmonic analysis), 2) meta-information (e.g., key and meter), 3) the context in which the piece was composed (e.g., history and story), and 4) subjective perceptions (e.g., sentiment and preference).

\section{Models}
Sequence-to-sequence generation tasks typically choose from three transformer-based architectures: encoder-decoder \cite{DBLP:conf/nips/VaswaniSPUJGKP17}, language model \cite{radford2018improving}, and prefix LM \cite{DBLP:conf/iclr/LiuSPGSKS18}. Based on previous findings \cite{DBLP:journals/jmlr/RaffelSRLNMZLL20}, we use the encoder-decoder architecture. As using pre-trained checkpoints can improve the performance of models in various tasks \cite{DBLP:journals/tacl/RotheNS20}, we hypothesised that it would be beneficial for the text-to-music generation task to use pre-trained checkpoints that already provide robust natural language representations. We use the following checkpoints to initialise the language-music model (see Table 1).

\begin{table}[t!]
  \begin{center}
    \caption{The configurations of various pre-trained checkpoints. The BART checkpoints initialise both the encoder and the decoder, while the rest of the checkpoints only initialise the encoder. The number of parameters in the encoder/decoder-only checkpoints varies slightly due to different sizes of vocabulary.}
    \begin{tabular}{l|c|c|c|c}
      \toprule 
      Checkpoint & Layers & Hidden & Heads & Params\\
      \midrule 
      RND & 12 & 768 & 12 & 91M\\
      BERT & 12 & 768 & 12 & 109M\\
      GPT-2 & 12 & 768 & 12 & 117M\\
      BART-base & 6+6 & 768 & 16 & 139M\\
      BART-large & 12+12 & 1024 & 16 & 406M\\
      \bottomrule 
    \end{tabular}
  \end{center}
\end{table}

\noindent
\textbf{RND (Random):} A randomly initialised encoder with a maximum input length of 1,024. We used byte-pair encoding for tokenization and only kept tokens with a minimum frequency of 100, ending up with a vocabulary size of 7,418. Setting a smaller minimum frequency would include a large number of unintelligible and meaningless tokens. 

\noindent
\textbf{BERT \cite{DBLP:conf/naacl/DevlinCLT19}:} It is an encoder-only bidirectional transformer pre-trained using a combination of masked language modelling objective and next sentence prediction on a large corpus comprising the Toronto Book Corpus and Wikipedia. We use the \texttt{bert-base-cased} checkpoint to initialise the encoder in our experiments.

\noindent
\textbf{GPT-2 \cite{radford2019language}:} It is a decoder-only, unidirectional transformer pre-trained using language modelling on a very large corpus of $\approx$40 GB of text data. We use the \texttt{gpt2-small} checkpoint to initialise the encoder (not the decoder) in our experiments.

\noindent
\textbf{BART \cite{DBLP:conf/acl/LewisLGGMLSZ20}:} It uses a standard encoder-decoder architecture with a bidirectional encoder (like BERT) and a unidirectional decoder (like GPT). The pre-training task involves randomly shuffling the order of the original sentences and a novel in-filling scheme, where spans of text are replaced with a single mask token. We use both checkpoints \texttt{bart-base} and \texttt{bart-large} in our experiments. The configurations of these two checkpoints do not exactly match those of others: the encoder/decoder of BART-base has only 6 layers (instead of 12), while BART-large has 1,024 units per layer (instead of 768), and they both have 16 heads (instead of 12). This is not compared apples to apples, but can provide us with baselines for initialisation using encoder-decoder checkpoints.

\begin{table*}[t]
  \begin{center}
    \caption{Results of various pre-trained checkpoints on the validation set. We found statistically significant improvements in BLUE-N and EDS for some pre-trained checkpoints, but not all of them had such benefits (e.g., BERT and BART-large).}
    \begin{tabular}{l|c|c|c|c|c|c|c}
      \toprule 
      Checkpoint & BLEU-2 & BLEU-3 & BLEU-4 & DIST-1 & DIST-2 & DIST-3 & EDS\\
      \midrule 
      RND & 44.47±20.64 & 35.88±18.12 & 28.84±16.22 & \textbf{10.61±4.32} & \textbf{28.47±11.03} & \textbf{41.79±15.24} & 38.02±13.10\\
      BERT & 21.72±14.01 & 15.44±10.61 & 10.47±7.90 & 8.71±5.49 & 19.03±11.95 & 27.09±17.03 & 24.81±10.40\\
      GPT-2 & 46.76±21.23 & 38.20±19.41 & 31.14±18.14 & 10.39±4.39 & 27.86±11.28 & 40.93±15.70 & 39.94±14.84\\
      BART-base & \textbf{48.34±21.47} & \textbf{39.85±20.11} & \textbf{32.82±19.22} & 10.29±4.31 & 27.98±11.01 & 41.28±15.15 & \textbf{40.77±15.75}\\
      BART-large & 22.41±14.28 & 15.96±10.71 & 10.91±7.78 & 9.21±6.36 & 20.54±13.28 & 29.09±18.04 & 24.98±10.27\\
      \bottomrule 
    \end{tabular}
  \end{center}
\end{table*}

Except for BART, the decoder for all other models is randomly initialised with the same configuration as the RND encoder. Since almost every character in the ABC notation is semantically independent, we took character-level tokenization (but added some common notations), with a vocabulary size of 164. We trained all models using the same learning rate $\alpha=10^{-4}$ (for BART-large, it is $5\times10^{-5}$), with a 1,000-step linear warmup and learning rate decay. We trained a total of 20 epochs with a batch size of 8, using the AdamW optimizer with $\beta_{1}=0.9$, $\beta_{2}=0.999$, $\epsilon=10^{-8}$, and a weight decay coefficient of 0.01.

\section{Experiments}
We randomly selected 2,828 (1\%) pairs from Textune as the validation set, and the rest were used for training. For both training and inference, we truncated all text sequences to the maximum input length of 1,024 (for BERT, it is 512). 

In Fig. 1, we display the training and validation curves for the five pre-trained checkpoints mentioned before. Note that the vocabulary size of the BART decoder (50,265 tokens) is much larger than that of the other randomly initialised decoders (164 tokens), which leads to higher losses but does not necessarily mean that the generation quality of BART is worse. Regardless, Fig. 1 suggests that using pre-trained checkpoints to initialise the model does not guarantee a lower validation loss.

Because of the small amount of data, all models showed different degrees of overfitting. In particular, even though the number of parameters is approximately three times that of BART-base, the validation loss of BART-large is not lower. Intuitive ways to solve this problem are to collect more data, reduce the model size, or tune hyperparameters. However, due to the scarcity of symbolic music data, it is unlikely to find a human-annotated text-music dataset that is at least an order of magnitude larger (i.e., 1 million text-music pairs) than Textune for a long time. Thus, using smaller models or tuning hyperparameters are attainable solutions for now.

To verify the generation quality, we used all checkpoints with their lowest validation loss to generate tunes based on descriptions from the validation set, and using nucleus sampling with top-$p=0.9$. We used the following metrics to evaluate the generated tunes from different models.

\textbf{BLEU-N \cite{papineni-etal-2002-bleu}:} An algorithm for evaluating the quality of text measures the proportion of N-grams in the reference text are reproduced by the candidate text. The higher the value, the closer the generated tunes are to ground truth. This is a common metric used in sequence-to-sequence tasks.

\textbf{DIST-N \cite{DBLP:conf/naacl/LiGBGD16}:} It evaluates the diversity of generated samples. A higher value of DIST-N means a higher proportion of distinct N-grams. We use this reference-free metric as text-to-music generation can be seen as conditional music generation, which is a creative task.

\begin{figure}[t]
	\centering
		\begin{minipage}{8.5cm} 
                        \includegraphics[width=\textwidth]{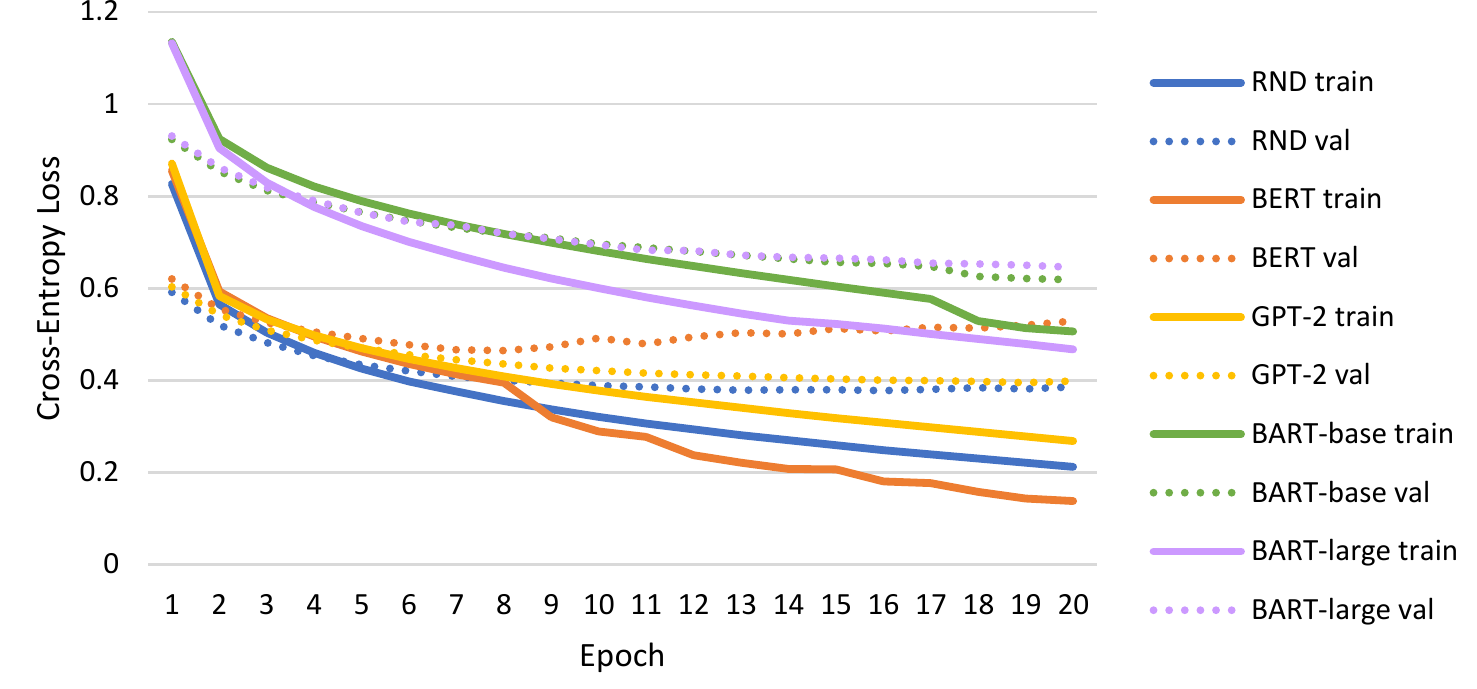}
		\end{minipage}
    \centering
	\caption{Training and validation curves of various models.} 
\end{figure}

\textbf{EDS:} \textbf{E}dit \textbf{D}istance \textbf{S}imilarity is based on the Levenshtein distance $lev(a,b)$ to indicate how similar the generated tune $b$ and the ground truth $a$ are at the character level, ranging from 0 (no match at all) to 100 (exact match), which can be formalised as follows:

\begin{equation}
    EDS(a,b) = (1 - \frac{lev(a,b)}{max(|a|,|b|)})\times 100,
\end{equation}
\noindent
where $|a|$ and $|b|$ are the length of two strings. As ABC tunes are nearly character-level sequences, EDS can effectively reflect the similarity between the generated tune and the ground truth.

As shown in Table 2, RND generated more diverse tunes (higher DIST-N), but the tunes generated by GPT-2 and BART-base are closer to the ground truth (higher BLEU-N and EDS). We performed independent samples $t$-tests, and found statistically significant differences in the BART-base results of BLEU-N and EDS compared to that of RND, i.e., $p$-value $<$ 0.05. These results show that the use of pre-trained checkpoints can improve the performance of the model on language-music tasks significantly.

For two randomly selected tunes from the Textune dataset, the average EDS is around 30\%, while the results for RND, GPT-2 and BART-base on this metric are around 40\%. Given the diversity of music, it indicates that these models can extract meaningful information from descriptions to generate music. However, the tunes generated by BERT and BART-large are not only low in diversity (lower DIST-N) but also far from the ground truth (lower BLEU-N and EDS). EDS suggests that the tunes generated by them are even more dissimilar to the ground truth compared to tunes randomly selected from Textune. We manually examined the tunes generated by these two models and found that there were a large number of instances of degeneration. We observed that they tend to repeatedly generate notes like \texttt{z8|z8|z8}. As shown in Fig. 1, we suggest that the cause of this problem is the severe overfitting of BERT and BART-large.

\begin{figure}[t]
    \centering
    \begin{minipage}{8.25cm} 
        \includegraphics[width=\textwidth]{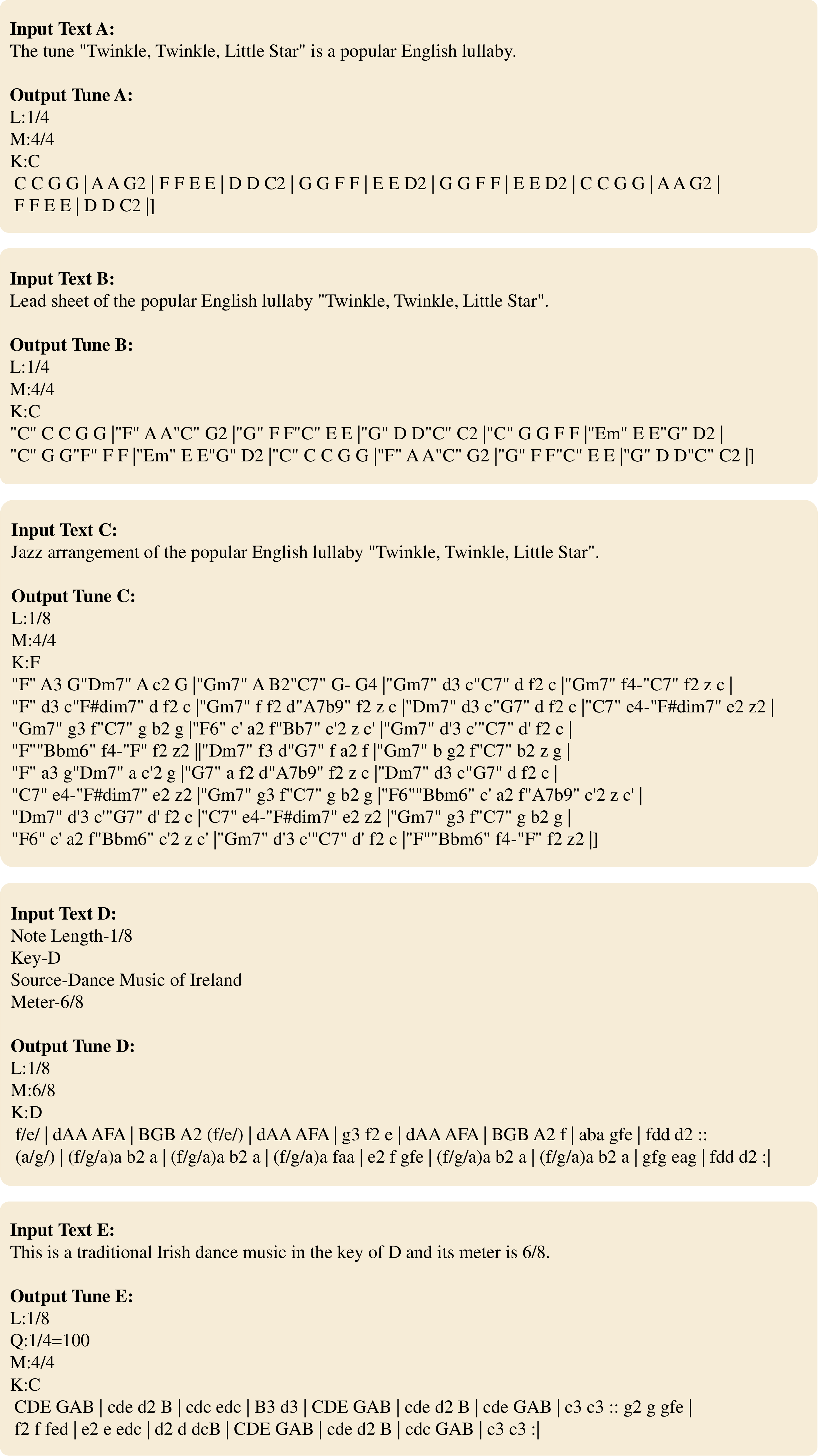}
    \end{minipage}
    \centering
    \caption{Music generation examples of RND} 
\end{figure}
\section{Discussions}
To demonstrate the capabilities and limitations of language-music models, several representative examples of generated tunes are given in Fig. 2. All the text descriptions in Fig. 2 were hand-crafted by us and are not from Textune. Due to space constraints, we only show RND-generated examples.

We first tested whether the model could reproduce the tunes already present in Textune. The tune chosen here is ``Twinkle, Twinkle, Little Star", which was present in Textune a total of 11 times. We found that when top-$p$ was set to a low value (e.g., 0.5), the model almost always reproduced the tune perfectly, as shown in Fig. 2A. This means that the model does understand the relationship between the title and the tune. However, this also indicates that it is possible for the model to directly copy the music that exists in Textune. We recommend using a higher top-$p$ when generating tunes using this model to avoid that problem.

We then tested the creativity of the model: generating the lead sheet and the jazz arrangement of ``Twinkle, Twinkle, Little Star". According to Fig. 2B, the model does understand what a lead sheet is and succeeds in placing appropriate chord symbols for this tune. It should be noted that Textune does not contain any lead sheets for this tune. This demonstrates the potential of language-music models to be applied to the melody harmonization task. However, for the more creative task, melody style transfer, the model did not perform well. The tune in Fig. 2C, although it does have a very distinctive jazz style (e.g., rhythm, harmony), has a completely different melody from  ``Twinkle, Twinkle, Little Star". Given that the model can perform well on the melody harmonization task, we believe that the reason for its failure on the melody style transfer comes mainly from the small amount of text-music data. If the size of text-music datasets can reach the level of text-image datasets \cite{DBLP:journals/corr/abs-2111-02114}, achieving most music generation tasks, including those requiring a high degree of creativity, should not be a challenge anymore.

We finally tested whether the model can follow the objective meta-information (e.g., key, meter) given in the text to generate tunes. We specified the key (D major), the meter (6/8), and the style of the music (Irish dance music). As shown in Fig. 2D and Fig. 2E, whether or not the model can generate music that matches the meta-information given in the text description depends on its format. When describing meta-information in a list format (Fig. 2D), the model can always follow the text accurately to generate tunes. The generated tune also exhibits distinctive characteristics of Irish dance music. For example, traditional Irish music is usually in a binary form (AABB), and the music generated here is exactly composed in that way. However, when the same information is given in a more loose way (Fig. 2E), the model does not follow the description well enough, even with a low top-$p=0.5$. Although the actual meter of this generated tune is still 6/8 and is in keeping with the characteristics of Irish music, the generated music is in the key of C major and the meter in the header is 4/4. We tested this text format on the dual task (i.e., music-to-text generation) and found that when given the prompt ``... in the key of", the model can always retrieve the meta-information correctly. Theoretically, the two tasks should be of equal difficulty, i.e., correctly translating the text to the header of ABC tunes or vice versa. More investigation is needed to determine the causes of this problem.

\section{Conclusions}
In this paper, we carry out the study of language-music models trained on large-scale text-music data. According to the experimental results, the use of pre-trained checkpoints leads to generated tunes that are much more similar to ground truth, but not improved in terms of diversity. Although the model can generate tunes that matched the semantic information of the text and exhibited a certain degree of creativity on some tasks, its creativity is limited, and it is input-sensitive. With a larger dataset, it is likely to develop a language-music model that performs well in music generation tasks that require a high degree of creativity.
\bibliography{aaai23}

\begin{thebibliography}{23}
\providecommand{\natexlab}[1]{#1}

\bibitem[{Briot(2021)}]{DBLP:journals/nca/Briot21}
Briot, J. 2021.
\newblock From artificial neural networks to deep learning for music
  generation: history, concepts and trends.
\newblock \emph{Neural Comput. Appl.}, 33(1): 39--65.

\bibitem[{Brown et~al.(2020)Brown, Mann, Ryder, Subbiah, Kaplan, Dhariwal,
  Neelakantan, Shyam, Sastry, Askell, Agarwal, Herbert{-}Voss, Krueger,
  Henighan, Child, Ramesh, Ziegler, Wu, Winter, Hesse, Chen, Sigler, Litwin,
  Gray, Chess, Clark, Berner, McCandlish, Radford, Sutskever, and
  Amodei}]{DBLP:conf/nips/BrownMRSKDNSSAA20}
Brown, T.~B.; Mann, B.; Ryder, N.; Subbiah, M.; Kaplan, J.; Dhariwal, P.;
  Neelakantan, A.; Shyam, P.; Sastry, G.; Askell, A.; Agarwal, S.;
  Herbert{-}Voss, A.; Krueger, G.; Henighan, T.; Child, R.; Ramesh, A.;
  Ziegler, D.~M.; Wu, J.; Winter, C.; Hesse, C.; Chen, M.; Sigler, E.; Litwin,
  M.; Gray, S.; Chess, B.; Clark, J.; Berner, C.; McCandlish, S.; Radford, A.;
  Sutskever, I.; and Amodei, D. 2020.
\newblock Language Models are Few-Shot Learners.
\newblock In \emph{Advances in Neural Information Processing Systems 33: Annual
  Conference on Neural Information Processing Systems 2020}.

\bibitem[{Chen et~al.(2021)Chen, Tworek, Jun, Yuan, de~Oliveira~Pinto, Kaplan,
  Edwards, Burda, Joseph, Brockman, Ray, Puri, Krueger, Petrov, Khlaaf, Sastry,
  Mishkin, Chan, Gray, Ryder, Pavlov, Power, Kaiser, Bavarian, Winter, Tillet,
  Such, Cummings, Plappert, Chantzis, Barnes, Herbert{-}Voss, Guss, Nichol,
  Paino, Tezak, Tang, Babuschkin, Balaji, Jain, Saunders, Hesse, Carr, Leike,
  Achiam, Misra, Morikawa, Radford, Knight, Brundage, Murati, Mayer, Welinder,
  McGrew, Amodei, McCandlish, Sutskever, and
  Zaremba}]{DBLP:journals/corr/abs-2107-03374}
Chen, M.; Tworek, J.; Jun, H.; Yuan, Q.; de~Oliveira~Pinto, H.~P.; Kaplan, J.;
  Edwards, H.; Burda, Y.; Joseph, N.; Brockman, G.; Ray, A.; Puri, R.; Krueger,
  G.; Petrov, M.; Khlaaf, H.; Sastry, G.; Mishkin, P.; Chan, B.; Gray, S.;
  Ryder, N.; Pavlov, M.; Power, A.; Kaiser, L.; Bavarian, M.; Winter, C.;
  Tillet, P.; Such, F.~P.; Cummings, D.; Plappert, M.; Chantzis, F.; Barnes,
  E.; Herbert{-}Voss, A.; Guss, W.~H.; Nichol, A.; Paino, A.; Tezak, N.; Tang,
  J.; Babuschkin, I.; Balaji, S.; Jain, S.; Saunders, W.; Hesse, C.; Carr,
  A.~N.; Leike, J.; Achiam, J.; Misra, V.; Morikawa, E.; Radford, A.; Knight,
  M.; Brundage, M.; Murati, M.; Mayer, K.; Welinder, P.; McGrew, B.; Amodei,
  D.; McCandlish, S.; Sutskever, I.; and Zaremba, W. 2021.
\newblock Evaluating Large Language Models Trained on Code.
\newblock \emph{CoRR}, abs/2107.03374.

\bibitem[{Davis and Mohammad(2014)}]{DBLP:conf/clfl/DavisM14}
Davis, H.; and Mohammad, S.~M. 2014.
\newblock Generating Music from Literature.
\newblock In \emph{Proceedings of the 3rd Workshop on Computational Linguistics
  for Literature, CLfL@EACL 2014}, 1--10.

\bibitem[{Devlin et~al.(2019)Devlin, Chang, Lee, and
  Toutanova}]{DBLP:conf/naacl/DevlinCLT19}
Devlin, J.; Chang, M.; Lee, K.; and Toutanova, K. 2019.
\newblock {BERT:} Pre-training of Deep Bidirectional Transformers for Language
  Understanding.
\newblock In \emph{Proceedings of the 2019 Conference of the North American
  Chapter of the Association for Computational Linguistics: Human Language
  Technologies, {NAACL-HLT} 2019}, 4171--4186.

\bibitem[{Dhariwal et~al.(2020)Dhariwal, Jun, Payne, Kim, Radford, and
  Sutskever}]{DBLP:journals/corr/abs-2005-00341}
Dhariwal, P.; Jun, H.; Payne, C.; Kim, J.~W.; Radford, A.; and Sutskever, I.
  2020.
\newblock Jukebox: {A} Generative Model for Music.
\newblock \emph{CoRR}, abs/2005.00341.

\bibitem[{He et~al.(2022)He, Chen, Xie, Li, Doll{\'{a}}r, and
  Girshick}]{DBLP:conf/cvpr/HeCXLDG22}
He, K.; Chen, X.; Xie, S.; Li, Y.; Doll{\'{a}}r, P.; and Girshick, R.~B. 2022.
\newblock Masked Autoencoders Are Scalable Vision Learners.
\newblock In \emph{{IEEE/CVF} Conference on Computer Vision and Pattern
  Recognition, {CVPR} 2022}, 15979--15988.

\bibitem[{Kreuk et~al.(2022)Kreuk, Synnaeve, Polyak, Singer, D{\'{e}}fossez,
  Copet, Parikh, Taigman, and Adi}]{DBLP:journals/corr/abs-2209-15352}
Kreuk, F.; Synnaeve, G.; Polyak, A.; Singer, U.; D{\'{e}}fossez, A.; Copet, J.;
  Parikh, D.; Taigman, Y.; and Adi, Y. 2022.
\newblock AudioGen: Textually Guided Audio Generation.
\newblock \emph{CoRR}, abs/2209.15352.

\bibitem[{Lewis et~al.(2020)Lewis, Liu, Goyal, Ghazvininejad, Mohamed, Levy,
  Stoyanov, and Zettlemoyer}]{DBLP:conf/acl/LewisLGGMLSZ20}
Lewis, M.; Liu, Y.; Goyal, N.; Ghazvininejad, M.; Mohamed, A.; Levy, O.;
  Stoyanov, V.; and Zettlemoyer, L. 2020.
\newblock {BART:} Denoising Sequence-to-Sequence Pre-training for Natural
  Language Generation, Translation, and Comprehension.
\newblock In \emph{Proceedings of the 58th Annual Meeting of the Association
  for Computational Linguistics, {ACL} 2020}, 7871--7880.

\bibitem[{Li et~al.(2016)Li, Galley, Brockett, Gao, and
  Dolan}]{DBLP:conf/naacl/LiGBGD16}
Li, J.; Galley, M.; Brockett, C.; Gao, J.; and Dolan, B. 2016.
\newblock A Diversity-Promoting Objective Function for Neural Conversation
  Models.
\newblock In \emph{{NAACL} {HLT} 2016, The 2016 Conference of the North
  American Chapter of the Association for Computational Linguistics: Human
  Language Technologies}, 110--119.

\bibitem[{Liu et~al.(2018)Liu, Saleh, Pot, Goodrich, Sepassi, Kaiser, and
  Shazeer}]{DBLP:conf/iclr/LiuSPGSKS18}
Liu, P.~J.; Saleh, M.; Pot, E.; Goodrich, B.; Sepassi, R.; Kaiser, L.; and
  Shazeer, N. 2018.
\newblock Generating Wikipedia by Summarizing Long Sequences.
\newblock In \emph{6th International Conference on Learning Representations,
  {ICLR} 2018}.

\bibitem[{Mohammad and Turney(2013)}]{DBLP:journals/ci/MohammadT13}
Mohammad, S.~M.; and Turney, P.~D. 2013.
\newblock Crowdsourcing a Word-Emotion Association Lexicon.
\newblock \emph{Comput. Intell.}, 29(3): 436--465.

\bibitem[{Papineni et~al.(2002)Papineni, Roukos, Ward, and
  Zhu}]{papineni-etal-2002-bleu}
Papineni, K.; Roukos, S.; Ward, T.; and Zhu, W.-J. 2002.
\newblock {B}leu: a Method for Automatic Evaluation of Machine Translation.
\newblock In \emph{Proceedings of the 40th Annual Meeting of the Association
  for Computational Linguistics}, 311--318.

\bibitem[{Radford et~al.(2018)Radford, Narasimhan, Salimans, Sutskever
  et~al.}]{radford2018improving}
Radford, A.; Narasimhan, K.; Salimans, T.; Sutskever, I.; et~al. 2018.
\newblock Improving language understanding by generative pre-training.

\bibitem[{Radford et~al.(2019)Radford, Wu, Child, Luan, Amodei, Sutskever
  et~al.}]{radford2019language}
Radford, A.; Wu, J.; Child, R.; Luan, D.; Amodei, D.; Sutskever, I.; et~al.
  2019.
\newblock Language models are unsupervised multitask learners.
\newblock \emph{OpenAI blog}, 1(8): 9.

\bibitem[{Raffel et~al.(2020)Raffel, Shazeer, Roberts, Lee, Narang, Matena,
  Zhou, Li, and Liu}]{DBLP:journals/jmlr/RaffelSRLNMZLL20}
Raffel, C.; Shazeer, N.; Roberts, A.; Lee, K.; Narang, S.; Matena, M.; Zhou,
  Y.; Li, W.; and Liu, P.~J. 2020.
\newblock Exploring the Limits of Transfer Learning with a Unified Text-to-Text
  Transformer.
\newblock \emph{J. Mach. Learn. Res.}, 21: 140:1--140:67.

\bibitem[{Ramesh et~al.(2022)Ramesh, Dhariwal, Nichol, Chu, and
  Chen}]{DBLP:journals/corr/abs-2204-06125}
Ramesh, A.; Dhariwal, P.; Nichol, A.; Chu, C.; and Chen, M. 2022.
\newblock Hierarchical Text-Conditional Image Generation with {CLIP} Latents.
\newblock \emph{CoRR}, abs/2204.06125.

\bibitem[{Rangarajan(2015)}]{DBLP:conf/icdim/Rangarajan15}
Rangarajan, R. 2015.
\newblock Generating music from natural language text.
\newblock In \emph{Tenth International Conference on Digital Information
  Management, {ICDIM} 2015}, 85--88.

\bibitem[{Reimers and Gurevych(2019)}]{DBLP:conf/emnlp/ReimersG19}
Reimers, N.; and Gurevych, I. 2019.
\newblock Sentence-BERT: Sentence Embeddings using Siamese BERT-Networks.
\newblock In Inui, K.; Jiang, J.; Ng, V.; and Wan, X., eds., \emph{Proceedings
  of the 2019 Conference on Empirical Methods in Natural Language Processing
  and the 9th International Joint Conference on Natural Language Processing,
  {EMNLP-IJCNLP} 2019}, 3980--3990.

\bibitem[{Rothe, Narayan, and Severyn(2020)}]{DBLP:journals/tacl/RotheNS20}
Rothe, S.; Narayan, S.; and Severyn, A. 2020.
\newblock Leveraging Pre-trained Checkpoints for Sequence Generation Tasks.
\newblock \emph{Trans. Assoc. Comput. Linguistics}, 8: 264--280.

\bibitem[{Schuhmann et~al.(2021)Schuhmann, Vencu, Beaumont, Kaczmarczyk,
  Mullis, Katta, Coombes, Jitsev, and
  Komatsuzaki}]{DBLP:journals/corr/abs-2111-02114}
Schuhmann, C.; Vencu, R.; Beaumont, R.; Kaczmarczyk, R.; Mullis, C.; Katta, A.;
  Coombes, T.; Jitsev, J.; and Komatsuzaki, A. 2021.
\newblock {LAION-400M:} Open Dataset of CLIP-Filtered 400 Million Image-Text
  Pairs.
\newblock \emph{CoRR}, abs/2111.02114.

\bibitem[{Vaswani et~al.(2017)Vaswani, Shazeer, Parmar, Uszkoreit, Jones,
  Gomez, Kaiser, and Polosukhin}]{DBLP:conf/nips/VaswaniSPUJGKP17}
Vaswani, A.; Shazeer, N.; Parmar, N.; Uszkoreit, J.; Jones, L.; Gomez, A.~N.;
  Kaiser, L.; and Polosukhin, I. 2017.
\newblock Attention is All you Need.
\newblock In \emph{Advances in Neural Information Processing Systems 30: Annual
  Conference on Neural Information Processing Systems 2017}, 5998--6008.

\bibitem[{Zhang et~al.(2020)Zhang, Wang, Wang, and Xia}]{zhang2020butter}
Zhang, Y.; Wang, Z.; Wang, D.; and Xia, G. 2020.
\newblock BUTTER: A representation learning framework for bi-directional
  music-sentence retrieval and generation.
\newblock In \emph{Proceedings of the 1st workshop on nlp for music and audio},
  54--58.

\end{thebibliography}
\end{document}